# Electrocapillary, thermocapillary, and buoyancy convection driven flows in Melcher – Taylor experimental setup

**Alexander Yu. Gelfgat\***

*School of Mechanical Engineering, Faculty of Engineering, Tel-Aviv University, Ramat Aviv, Tel-Aviv 6997801, Israel, gelfgat@tau.ac.il*

**Gerrit Maik Horstmann**

*Helmholtz-Zentrum Dresden - Rossendorf, Bautzner Landstr. 400, 01328 Dresden, Germany, g.horstmann@hzdr.de*

Abstract

Electrocapillary-driven two-phase flows in a confined configuration of a classical experiment of Melcher and Taylor are studied. The computed streamlines of the flow of the heavier dielectric liquid (corn oil) qualitatively represents the corresponding experimental image. With the increase of electrocapillary forcing, the flow pattern changes, so that the main circulation localizes near boundary with a larger electric potential. When a dielectric liquid is replaced by a poorly conducting one, the system becomes non-isothermal owing to the Joule heating. Then the flow is driven also by buoyancy and thermocapillary convection, whose effect becomes noticeably stronger than the electrocapillary one. With the increase of electric conductivity, the electrocapillary effect is further weakened compared to the two others, while the electrocapillary and thermocapillary forces remain comparable at the central part of the interface, where the thermocapillary force changes its sign. The results show that consideration of the two-phase model is mandatory for obtaining correct flow patterns in the lower heavier fluid. The Lippmann equation, connecting electrically induced surface tension with non-uniform surface electric potential, is numerically verified for both isothermal and non-isothermal formulations and was found to hold in both of them.

1. <u>Introduction</u>

Electrocapillary flows appear in two-liquid systems under action of an external electric field tangent to the liquid-liquid interface. An excessive surface charge, which can result either from charge carrying particles or polarization of electrically neutral ones, interacts with the electric field, which creates a Coulomb force that drives the flow. This force is located at the interface and sometimes is interpreted as an electrostatic addition to the surface tension.

Melcher and Taylor [1] observed some earlier works, gave a general description of the phenomena of electrocapillarity, and presented several examples of electrocapillary flows. Later studies considered mainly flows of drops and small particles [2-3], thin films and sheets [4-8], electrocapillary instability of quiescent [9-12] and parallel [14-17] flow in two-fluid systems, and electrocapillary flows in bounded cylindrical geometries [18-22]. One part of the studies was focused on deformations of the liquid-liquid interface [4-8,22], while the other part considered instabilities of quiescent fluid or simple shear flows under effect of the electrocapillary forces [14-17,21]. Analytical solutions for developed electrocapillary driven flows are given in [1] for an infinite layer and a Stokes flow around a drop.

Comparing to an extensive literature on similar thermocapillary flows also driven by the surface force (see [23] and references therein), there is a certain lack of results for developed non-linear electrocapillary flows in finite, experimentally realizable, geometry. To the best of our knowledge, the study [24] was the only attempt to approach numerically the model experiment of Melcher and Taylor [1]. The authors of [24] reported flow patterns qualitatively similar to those observed in the experiment, however, values of their dimensionless parameters did not exactly correspond to the parameters of the experiment.

This study consists of two main parts. In the first one we address numerically the model experiment of [1] that illustrated the electrocapillary driving effect in a rectangular cavity filled with a dielectric liquid and covered by a triangular air layer. We formulate a two-phase two-dimensional numerical model, which allows us to represent qualitatively the experimental observations. Further computational modeling shows that with increase of the voltage the electrocapillary flow pattern changes similarly to thermocapillary flow driven by an increasing temperature gradient along the interface [23,56]. In both cases, the main circulation becomes more intensive either near a cold vertical wall, or near the wall with a lower electric potential.

The second part is motivated by electrocapillary phenomena observed in liquid electrolytes, in particular, liquid electrodes [24] in liquid metal batteries (LMBs) [26]. We argue that a similar Taylor-Melcher setup [1], where the dielectric fluid is replaced by a poorly conducting one, will be subject to Joule heating. The resulting non-uniform distribution of the temperature will trigger buoyancy and thermocapillary convection. We find that with the increase of electric conductivity, the electrocapillary forcing becomes small compared to the two convective forces. The results show also that consideration of the two-phase model, and taking buoyancy convection in the air into account, are crucial for obtaining a correct flow pattern in the lower heavier liquid.

Finally, we examine whether obtained numerical solutions obey the Lippmann equation [27], which describes dependence of the surface tension on the applied voltage. The results show that the Lippmann equation is perfectly satisfied both for isothermal and non-isothermal cases. Since this equation is derived on the basis of purely thermodynamic considerations, which are not explicitly included in our numerical model, we consider this test as an independent validation of the results.

## 2. Surface force induced by electrocapillarity

Since electrocapillary flows are much less known compared to thermocapillary or concentration-capillary flows, we give a brief introduction to the basic physical principles. The field of Electrohydrodynamics treats flows induced by electric fields in weakly-conducting liquids such as leaky dielectrics or poorly conducting fluids, e.g., aqueous electrolytes [27-33]. The electromechanical coupling at the interfaces, where electrical parameters are subject to discontinuities, induce unneglectable surface forces that must be accounted for in fluid dynamics models. The effect of electrocapillarity relates to variation of surface tension $\gamma$ with the local interfacial electrical potential $V$, similarly to better known variations of the surface tension with temperature $T$ or concentration $c$. The dependence $\gamma(T,c)$ is a material property, and usually is assumed to be linear for both the temperature and the concentration. In most liquids $\partial\gamma/\partial T < 0$, which defines the direction of the thermocapillary force from warmer to colder interface regions. In the electrocapillary flows dependence of the surface tension on the local electric potential $V$ is not monotonic, usually parabolic $\gamma \sim V^2$, so that the sign of $\partial\gamma/\partial V$ can change. The parabolic dependence follows from the Lippmann equation [34,35], which, under the constraint that liquid composition and temperature remain quasi-constant, states

$$\frac{\partial \gamma}{\partial V} = -q_s \; , \tag{1}$$

where $q_s$ is the interface electric charge. The latter can be either induced by external electric fields, or result from covalently bound ionizable groups, or ion adsorption in electrochemical systems [32,35,36]. However, the charge density is not a material constant, but results from electrohydrodynamics of the whole system. To avoid consideration of the full electrohydrodynamic problem, the interface can be assigned with a specific capacitance $C$, such that $q_s = C(V - V_0)$, where $V_0$ denotes the potential of zero charge, which is nonzero in most electrochemical systems. In the particular case of the dielectric liquid addressed in this paper, an external potential difference is needed to induce surface charges, and the capacitance is simply given by the electric permittivity and the normal distance between the electrodes, $C = \varepsilon_0 / a$ (see below). Substituting $q_s = C(V - V_0)$ into equation (1) and integrating yields the parabolic electrocapillary profile

$$\gamma = \gamma_0 - \frac{1}{2} C(V - V_0)^2, \tag{2}$$

with $\gamma_0 = \gamma(V = V_0)$ being the surface tension in the discharged case. Equation (1) shows that any increase in charge density, positive or negative, will result in a decrease in the surface tension. This behavior can be intuitively explained by the fact that all interfacial charges exert repulsive Coulomb forces on each other, which are tangential to the interface. A hypothetical expansion of the interface therefore requires less energy resulting in lower surface tension. Electrocapillary flows necessarily develop whenever there is an unbalanced tangential gradient of the surface tension created by tangential voltage gradient, which, in its turn, results from a nonzero tangential electric field component. Contrarily to thermocapillary flows the derivative $\partial \gamma / \partial V$ changes with polarization ($V < V_0$ or $V > V_0$), which makes electrocapillary forcing more complicated, potentially exhibiting both destabilizing or stabilizing effects. Apparently, to attain sufficiently large tangential stresses that can noticeably affect macroscopic flows, either the capacitance or the voltage must be large enough. In dielectrics, the capacitance is very small such that voltages in the order of kilovolts must be applied to drive considerable flows. In contrast, the capacitance of electrochemical double layers is very high, allowing to reach similar tangential stresses only with a few volts. The latter is illustrated by the results below. As a last point, it should be mentioned that the introduced electrocapillary effect can also be described and understood in a fully equivalent way by considering electric Maxwell stresses acting at the interface [27,29]. This formulation is applied in this study, while the Lippmann equation (1) is used for validation of the numerical results.

## 3. Description of the problem

We consider a 2D configuration of the experiment described in Melcher & Taylor [1], see their Figs. 1 and 2. The problem is sketched again in Fig. 1 below. The liquid (corn oil in [1]) occupies a rectangular volume of length $l$ and height $b$. The vertical boundaries are electrodes that maintain the electric potential difference $V_0$. The third electrode is a straight plate, whose left end is attached to the top of the left vertical boundary, and the right end is positioned at the distance $a$ above the top of the right boundary. Its electric potential is the same as that of the left boundary. The bottom and part of the vertical boundary that connects the second and the third electrodes are assumed to be electrically insulated. The triangular part above the liquid is filled with air.

The jump of the electrical field vertical component over the liquid-air interface creates an excessive surface charge, whose interaction with the tangential component of electric field results in the Coulomb force acting along the surface, as explained above and in [1]. We assume that the gravity force is large enough to keep the interface flat, which is consistent with the experimental photo in Fig. 2 of [1]. More arguments justifying this assumption are given below. The horizontal component of the Coulomb force acts along the interface and drives the flow. The whole system represents a two-phase flow driven by the tangential Coulomb force acting along the interface. In all the calculations below, the upper fluid is air and the upper boundary is at room temperature. The lower and heavier fluid is taken first as a dielectric corn oil (liquid 1) used in the experiment [1]. Then we examine the electrocapillary driving in poorly conducting liquids with gradually increasing electric conductivity. For representative examples we take 0.1 Mole/kg NaCl water solution (liquid 2), 2 Mole/kg NaCl solution (liquid 3), and LiCl-KCl mixture used in some liquid metal batteries [30,31]. For further clearness we collect all the material properties used in the following calculations in Table 1, and the corresponding dimensionless numbers in Table 2. The properties were taken from the published data [39-55].

### 5. Governing equations

We use the leaky dielectric model introduced in [1] and thoroughly re-derived by Saville [27]. Since the electric field is irrotational, it is represented by the gradient of the electric potential $\boldsymbol{E} = -\nabla\phi$. Neglecting the volume electric charges in each of the phases 1 and 2, continuity of the electric current yields

$$div(\sigma_j \boldsymbol{E}_j) = 0 \implies \Delta\phi_j = 0, \quad j = 1,2. \tag{3}$$

The boundary conditions for the electric potential follow from the description of Fig. 1:

$$\phi_1(x = 0, y) = \phi_2\left(x, y = b + a\frac{x}{l}\right) = 0, \tag{3.1}$$

$$\phi_1(x = l, 0 \leq y \leq b) = V_0, \tag{3.2}$$

$$\frac{\partial\phi_1}{\partial y}(x, y = 0) = \frac{\partial\phi_2}{\partial x}(x = l, b < y \leq b + a) = 0. \tag{3.3}$$

At the interface $y = b$:

$$\phi_1(x, y = b) = \phi_2(x, y = b), \tag{4.1}$$

$$\varepsilon_0\left[\varepsilon_2 E_{2,y} - \varepsilon_1 E_{1,y}\right]_{y=b} = \varepsilon_0\left[\left(\varepsilon_1\frac{\partial\phi_1}{\partial y} - \varepsilon_2\frac{\partial\phi_2}{\partial y}\right)\right]_{y=b} = q_s \,, \tag{4.2}$$

where $q_s$ is the surface charge, $\varepsilon_0$ is electric permittivity of vacuum, and $\varepsilon_j$ are electric permittivities of the two liquids. To close formulation for the electric field, we need to add an equation describing conservation of the surface charge, which will include the flow velocity and will be discussed later.

In the considered model, the electrocapillary flow appears after the voltage $V_0$ is applied (see Fig. 1), which leads to a non-uniform distribution of the electric charge along the interface. At the same time, assuming that the boundaries with different electric potential are connected via a closed circuit, the electric current will lead to release of Joule heat in the lower poorly conducting liquid. This can be significant even for fluids with very small electrical conductivity, like it happens, e.g., in liquid metal batteries [30]. Effect of the Joule heating is two-fold. First, a non-uniform distribution of the temperature triggers natural convection. Second, a temperature variation along the interface creates thermocapillary forces that will interact with the electrocapillary ones. Taking all these into account, and neglecting all possible effects of the flow on the imposed electric field, the remaining part of the problem is described by the momentum, continuity, and internal energy equations. For further formulation, it is necessary to discuss the electrostriction force, which is derived from the Maxwell stress tensor [32,33], and must be added to the momentum equation. The general expression for this force is [32]

$$f_i^E = -\frac{1}{2}\frac{\partial}{\partial x_i}\left[\left(\varepsilon - \rho\frac{\partial \varepsilon}{\partial \rho}\right)\boldsymbol{E}^2\right] + \frac{\partial}{\partial x_j}\left(\varepsilon E_i E_j\right) \; . \tag{5.1}$$

The first term of the above expression is potential and can be added to the pressure gradient term of the momentum equation. This term does not affect the velocity field, but should be taken into account for derivation of the shape of the interface. The second term can be expressed as $Div[\varepsilon \boldsymbol{E} \otimes \boldsymbol{E}]$ and can be evaluated using the relation $Div[\boldsymbol{U} \otimes \boldsymbol{W}] = \boldsymbol{U} div \boldsymbol{W} + (\boldsymbol{W} \cdot \nabla)\boldsymbol{U}$. Assuming $\varepsilon = const$ in each layer,

$$Div[\varepsilon \boldsymbol{E} \otimes \boldsymbol{E}] = \varepsilon \boldsymbol{E} div \boldsymbol{E} + \varepsilon (\boldsymbol{E} \cdot \nabla)\boldsymbol{E} = \varepsilon \boldsymbol{E} div \boldsymbol{E} + \frac{1}{2}\varepsilon \text{grad}(\boldsymbol{E}^2) - \varepsilon \boldsymbol{E} \times rot \boldsymbol{E} \tag{5.2}$$

The first term of Eq. (5.2) together with the Maxwell equation $\varepsilon div \boldsymbol{E} = q$, yields the Coulomb force $\boldsymbol{f}^C = q\boldsymbol{E}$ , where $q$ is the volume charge. Since there is no free charge in the considered system, the first term of r.h.s. of Eq. (5.2) can be dropped. According to the Maxwell equation $rot \boldsymbol{E} = -\partial \boldsymbol{B}/\partial t$, and $rot \boldsymbol{E} = \boldsymbol{0}$ for the steady electromagnetic field, so that the third term also vanishes. The second term is potential and can be added to the pressure gradient term. Assuming constant electric permittivity, the resulting electrostriction pressure results from Eq. (5.1)

$$p_{str} = \frac{1}{2}\varepsilon \boldsymbol{E}^2, \quad \boldsymbol{f}_{str} = \nabla p_{str} \; . \tag{5.3}.$$

To apply the Boussinesq approximation, we assume a linear dependence of the densities on the temperature, $\rho_j = \rho_{0,j}[1 - \beta_j(T_j - T_0)]$, where $\beta_j$ are the thermal expansion coefficients of the liquid and the air, and $\beta_j(T_{max} - T_{min}) \ll 1$. Governing equations for velocity $\boldsymbol{u}_j = (u_j, w_j)$, pressure $p_j$, and temperature $T_j$, $j =$1 or 2, of the liquids 1 and 2, read

$$\rho_j\left[\frac{\partial \boldsymbol{u}_j}{\partial t} + (\boldsymbol{u}_j \cdot \nabla)\boldsymbol{u}_j\right] = -\nabla P_j + \mu_j \Delta \boldsymbol{u}_j + g\beta_j(T_j - T_0)\boldsymbol{e}_y, \tag{6}$$

$$\nabla \cdot \boldsymbol{u}_j = 0, \tag{7}$$

$$\rho_j c_{p,j}\left[\frac{\partial T_j}{\partial t} + (\boldsymbol{u}_j \cdot \nabla)T_j\right] = \kappa_j \Delta T_j + \frac{J_j^2}{\sigma_j}, \tag{8}$$

Here $t$ is time, $\mu_j$, $T_j$, $c_{p,j}$, $\kappa_j$, and $\sigma_j$ are viscosity, temperature, heat capacity, heat conductivity, and electric conductivity of the fluids, respectively. $\boldsymbol{J}_j = \sigma_j \nabla \phi_j$ is the electric current in each layer. The total pressure $P_j$ consists of dynamic, hydrostatic and electrostriction parts, and is expressed as

$$P_j = p_j + \rho_{0,j} g + \tfrac{1}{2} \varepsilon_j \boldsymbol{E}_j^2 \qquad (9)$$

Assuming all the boundaries to be no-slip and isothermal at the temperature $T_0$, the boundary conditions read

$$\boldsymbol{u}_j(x, y=0) = \boldsymbol{u}_j(x=0, y) = \boldsymbol{u}_j(x=l, y) = \boldsymbol{u}_j\left(x, y = b + a\frac{x}{l}\right) = 0, \qquad (10)$$

$$T_j(x, y=0) = T_j(x=0, y) = T_j(x=l, y) = T_j\left(x, y = b + a\frac{x}{l}\right) = T_0, \qquad (11.1)$$

$$\kappa_1 \left(\frac{\partial T_1}{\partial y}\right)_{y=b} = \kappa_2 \left(\frac{\partial T_2}{\partial y}\right)_{y=b}. \qquad (11.2)$$

To render equations dimensionless, we choose $b$ to be the length scale, and, following common formulations for natural convection, the velocity scale is defined as $\mu_1/\rho_1 b$. The scales of time, pressure, electric potential and electric current are $b^2\rho_1/\mu_1$, $\mu_1^2/b^2\rho_1$, $V_0$, and $\sigma_{0,m}V_0/b$, respectively. The temperature is rendered dimensionless by $\theta = (T - T_0)/(T_1 - T_0)$, and assuming the Boussinesq approximation, which includes Eq. (9), we arrive to the non-dimensional form of equations

$$\left[\frac{\partial \boldsymbol{u}_j}{\partial t} + \left(\boldsymbol{u}_j \cdot \nabla\right)\boldsymbol{u}_j\right] = -\frac{\rho_{2,0}}{\rho_{21}\rho_{j,0}}\nabla P_j + \mu_j \frac{\rho_{2,0}}{\rho_{21}\rho_{j,0}}\frac{\mu_{21}\mu_j}{\mu_2}\Delta \boldsymbol{u}_j + Gr\frac{\beta_{21}\beta_j}{\beta_2}\theta_j \boldsymbol{e}_y, \qquad (12)$$

$$\nabla \cdot \boldsymbol{u}_j = 0, \qquad (13)$$

$$\frac{\partial T_j}{\partial t} + \left(\boldsymbol{u}_j \cdot \nabla\right)\theta_j = \frac{\alpha_{21}\alpha_j}{\alpha_2}\frac{1}{Pr}\Delta \theta_j + \frac{\rho_{2,0}}{\rho_{21}\rho_{j,0}}\frac{c_{p,2}}{c_{p,21}c_{p,j}}\frac{\sigma_2}{\sigma_{21}\sigma_j}Jo\boldsymbol{J}^2, \quad \boldsymbol{J}_j = \frac{\sigma_{21}\sigma_j}{\sigma_2}\nabla \phi_j, \qquad (14)$$

where the dimensionless governing parameters are ratios of physical properties $\rho_{21} = \rho_2/\rho_1$, $\mu_{21} = \mu_2/\mu_1$, $\beta_{21} = \beta_2/\beta_1$, $\sigma_{21} = \sigma_2/\sigma_1$, $\varepsilon_{21} = \varepsilon/\varepsilon_1$ and $\alpha_{21} = \alpha_2/\alpha_1$ $(\alpha_j = k_j/\rho_j C_{p,j})$, the Prandtl number $Pr = \mu_1/\alpha_1\rho_1$, the Grashof number $Gr = g\beta_1(T_1 - T_0)H^3\rho_{1,0}^2/\mu_1^2$, and the Joule heating number $Jo = \sigma V_0^2 b^2/\mu_1 c_{p,1}(T_1 - T_0)$.

The shape of the liquid-liquid interface is described as $y/b = 1 + \chi(x)$. It is defined by the balance of normal stresses, which, in the dimensionless form reads

$$p_1 - p_2 + \tfrac{1}{2}Ec(E_1^2 - \varepsilon_{21}E_2^2) - Ga\left(1 - \frac{1}{\rho_{21}}\right)\chi - \frac{Ga}{Bo}K = -\left[\frac{\partial u_1}{\partial z} - \mu_{21}\frac{\partial u_2}{\partial z}\right]n_i n_k, \qquad (15)$$

where $K$ is the dimensionless main interface curvature, $\boldsymbol{n}$ and $\boldsymbol{\tau}$ are unit normal and tangent to the interface vectors, respectively, $Ga = \frac{g\rho_{1,0}^2 b^3}{\mu_1^2}$ is the Galileo number, $Bo = \rho_{1,0} g b^2/\sigma$ is the Bond number, and $Ec = \frac{\varepsilon_1 V_0^2 \rho_1}{\mu_1^2}$ is a dimensionless parameter that describes the electrocapillary forcing and sometimes called electrocapillary Marangoni number. As follows from Table 2, the Galileo number $Ga$ is several orders of magnitude larger than the electrocapillary number $Ec$, as well as than the ratio $Ga/Bo$. The other terms are noticeably smaller, which can be seen, for example, from the numerical results. This means that all the terms of Eq. (15) can be balanced only if the interface deviation from the horizon $\chi$ and the curvature $K$ are several orders smaller than unity, meaning that the interface deformations can be neglected. The results of [24], where these deformations were included in the model, did not report any significant interface deviation from the flat shape. It is stressed, however, that investigation of stability of steady flows reported below requires consideration of the interface deformations, as is argued in [27].

For the flat interface, the normal to interface velocity is zero,

$$u_y(x, y = b) = 0. \qquad (16)$$

The tangent stress balance at the flat interface reads

$$\left[\mu_2 \frac{\partial u_{2,x}}{\partial y} - \mu_1 \frac{\partial u_{1,x}}{\partial y}\right]_{y=b} = q_s E_x(x, y = b) + \frac{d\gamma}{dx} = q_s \left[\frac{\partial \phi}{\partial x}\right]_{y=b} - \left|\frac{d\gamma}{dT}\right|\frac{dT}{dx}, \qquad (17)$$

Following the introduced above scales, we define scales of the electric field and the surface charge as $V_0/b$ and $\varepsilon_0 V_0/b$, respectively. This yields Eq. (17) in the dimensionless form

$$\left[\mu_{21} \frac{\partial u_{2,x}}{\partial y} - \frac{\partial u_{1,x}}{\partial y}\right]_{y=b} = Ec\left[\frac{\partial \phi}{\partial x}\right]_{y=b} - MaPr\frac{dT}{dx}, \qquad (18)$$

where $Ma = \left|\frac{d\gamma}{dT}\right|\frac{(T_1-T_0)b}{\mu_1 \alpha_1}$ is the Marangoni number.

In the work of Saville [27] the conservation of surface charge is described by equation (22'), in which the charge diffusion is neglected. Also, the origin of the velocity dependent terms is not completely clear. In our opinion, keeping the diffusion term is essential, since otherwise it will be impossible to arrive to a steady distribution of the surface charge in a quiescent fluid. In the following we use the equation derived in [36] and used in [15], which reads for an arbitrary interface

$$\frac{\partial q_s}{\partial t} + u_n(\nabla \cdot \boldsymbol{n})q_s + \nabla_s \cdot \boldsymbol{K}_s + \boldsymbol{n} \cdot [\![\sigma \boldsymbol{E}]\!] - u_n[\![q]\!] = 0, \qquad (19)$$

where

$$K_s = \xi_s q_s \boldsymbol{E}_s + \sigma_s \boldsymbol{E}_s + q_s \boldsymbol{u}_s - D_s \nabla q_s. \tag{20}$$

Here $\boldsymbol{n}$ is the normal to the interface, $\xi_s$ is the surface electrical mobility of ions, $\sigma_s$ is interface electrical conductivity, $D_s$ is the surface diffusion coefficient, $u_n$ is the normal to the surface velocity, and $(\nabla \cdot \boldsymbol{n})$ represents the interface curvature. The operator $\nabla_s$ is the projection of the operator $\nabla$ on the interface and is defined as $\nabla_s = \nabla - \boldsymbol{n}(\boldsymbol{n} \cdot \nabla)$. The double square brackets denote jump of a quantity from phase 1 to phase 2, $[\quad]_2 - [\quad]_1$. A similar formulation can be found in [32,33].

For the present problem $u_n = 0$, $(\nabla \cdot \boldsymbol{n}) = 0$, and $\nabla_s = \partial/\partial x$. Therefore, for a steady flow, Eqs. (9) and (10) yield

$$\xi_s \frac{\partial}{\partial x}(q_s E_x) + \sigma_s \frac{\partial E_x}{\partial x} + q_s u_x - D_s \frac{\partial^2 q_s}{\partial x^2} + \left(\sigma_1 E_{1,y} - \sigma_2 E_{2,y}\right) = 0 \quad . \tag{21}$$

Considering boundary conditions for the surface charge, we assume that there is no flux of the surface charge through the left boundary. For the right boundary this assumption leads to a discontinuity in the derivative $\frac{\partial q_s}{\partial y}$ at $y = b$. Thus, we derive the boundary condition from Eq. (4.2) taking into account that this boundary is equipotential for $y \leq b$, so that

$$\left[\frac{\partial q}{\partial x}\right]_{y=b, x=0} = 0, \quad q_s(x = l) = \varepsilon_0 \varepsilon_2 \left[\frac{\partial \phi_2}{\partial y}\right]_{y=b, x=l}, \tag{22}$$

This closes the set of equation for the considered problem.

The problem is solved using the finite volume discretization on a regular uniform grid. The upper boundary is accounted for by the immersed boundary method. The steady flows are calculated by the Newton method using the approach of [56]. The calculations are performed on grids varying from 200×200 to 500×500 nodes, stretched as in [56] in the cases where boundary layers observed. The convergence study shows that this grid yields three decimal places converged for all computed functions.

## 6. Plane-parallel solution of Melcher and Taylor.

Melcher & Taylor [1] considered the current problem assuming very large length $l$ and vacuum as the second phase. The electric potential was assumed to be a linear function of $x$ and $y$, so that the horizontal and vertical components of the electric field were constant:

$$\phi = V_0 \left(\frac{x}{l} - \frac{y}{a}\right) , \qquad E_x = -\frac{\partial \phi}{\partial x} = -\frac{V_0}{l} \qquad , \qquad E_y = -\frac{\partial \phi}{\partial y} = \frac{V_0}{a}. \tag{23}$$

This yields for the electric charge and the Coulomb tangent stress

$$q_s = \frac{\varepsilon_0 V_0}{a}, \qquad \tau_C = -\varepsilon_0 \frac{V_0^2}{al} \quad . \tag{24}$$

Now assuming that owing to a large $l$, velocity has only horizontal non-zero component that depends only on the coordinate $y$, the problem for velocity becomes

$$\mu \frac{d^2 u_x}{dy^2} = -\frac{dp}{dx}, \qquad u_x(y=0) = 0, \qquad \mu \left[\frac{du_x}{dy}\right]_{y=b} = -\varepsilon_0 \frac{V_0^2}{al} \tag{25}$$

The pressure gradient $\frac{dp}{dx}$ is assumed to be a constant, and is obtained from the requirement of the zero mass flus across the liquid layer

$$\int_0^b u_x dy = 0 \tag{26}$$

The solution of (16), (17) is

$$u_x = -\frac{\varepsilon_0}{2\mu} \frac{V_0^2 b}{al} \left(\frac{3}{2}\frac{y^2}{b^2} - \frac{y}{b}\right) \tag{27}$$

This is the same parabolic return flow, as the profile of Birikh [57] for thermocapillary convection in a horizontal layer subject to a horizontal temperature gradient.

7. <u>Flow in a finite geometry</u>

As mentioned above, for calculations in the finite geometry that fully corresponds to the geometry of experiment [1], we consider four different dielectric and poorly conducting liquids, whose properties taken from [39-55] are listed in Table 1. First we consider the corn oil (liquid 1) used in the above experiment. Then, we are interested in the effect of release of the Joule heat when the lower liquid is weakly electrically conducting. For this purpose, we consider a dilute 0.1 Mole/kg NaCl-water solution (liquid 2), a more concentrated 2 Mole/kg NaCl-water solution (liquid 3), whose electric conductivity is noticeably larger, and, finally, mixture of LiCl-KCl salts (liquid 4), whose properties are taken as characteristic for electrolytes used in the LMB design [30,31], and whose electric conductivity is larger than that of the two previous liquids. Note that the values of surface conductivity, mobility, and diffusion coefficient are unknown, and we estimate them by the order of magnitudes of these values published for other materials.

To gain better understanding of results of the Joule heating and magnitudes of flow velocities, all the results are reported as dimensional values. The dimensionless parameters, which allow for a better comparison between the cases, are listed in Table 2.

### 7.1. The corn oil – air system (Melcher-Taylor experiment [1])

The experimental photo of the flow pattern is reproduced in Fig. 2a. To compare, we first perform calculations considering the same finite geometry and voltage, but a constant interface Coulomb tension given by Eq. (23). The resulting streamlines are plotted in Fig. 2b. Then, we perform computations for the full model (3)-(22), which results in the streamlines plotted in Fig. 2c. We observe that the flows are similar, and the minimal and maximal values of the stream function are close. The flow along the interface is directed from the right to the left boundary, i.e., from the larger to lower voltage. The maximum of stream function is shifted towards the left boundary, which is consistent with the experimental photo of [1] shown in Fig. 2a. This is similar to the patterns developing in long horizontal cavities under action of thermocapillary force [37]. Note, that the interface stress drives also the air flow in the upper triangular part of the setup. Owing to the small viscosity of air, its influence on the flow inside the oil is negligible. However, its effect can be more significant if, say, the two liquids are non-isothermal and continuity of heat flux at the interface is required (see below).

Similarities and differences in the description of the electric part in the two cases are illustrated in Fig. 3. Figure 3a shows equipotential lines for the full leaky dielectric model. We see that the potential inside the oil is a linear function of $x$, and inside the air it is almost a linear function of $x$ and $y$, as it was assumed by Melcher & Taylor [1]. The difference is clearly seen at the profile of surface charge (Fig. 3b). The surface charge is almost constant, except a short region near the right boundary, where it steeply grows. The value of constant is $q_s \approx 6.4 \cdot 10^{-6} \, C/m^2$, while Eq. (15) yields a close value $q_s \approx 6.2 \cdot 10^{-6} \, C/m^2$. The steep growth near the right boundary and a slight decay at the left boundary are caused mainly by changes of $\partial \phi / \partial y$ at the air side of the interface.

To look for possible flow changes, we varied the last three parameters of Table 1, which are unknown for the corn oil – air system. Increase of the electric mobility to the value of $10^{-3} \frac{m^2}{V \cdot s}$, or surface diffusion coefficient to $0.01 \frac{m^2}{s}$, or surface conductivity to $10^{-5} \, 1/\Omega \cdot m^2$, did not noticeably alter the whole numerical solution. This shows that added phenomena of surface charges mobility, surface diffusion, and surface conductivity are negligible compared to the effect of the surface charge accumulation owing to the electric field inhomogeneity. It points also on the consistency of the Melcher-Taylor model. At the same time, these meanwhile negligible effects may affect stability of the interface, which is yet to be studied.

A possible change takes place with further increase of the applied voltage. This is illustrated in Fig. 4. We observe that with the increase of voltage, meaning increase of the forcing, the main circulation is shifted towards the left boundary, where the most intensive flow is localized. Again, this is similar to changes of flow patterns of thermocapillary convection with the increase of the Marangoni number [56,37].

Finally, we verify additionally the electrostatic part of our numerical model. The additional surface tension induced by the Coulomb forces is characterized by the surface tension coefficient $\gamma$, which is analogous to the mechanical surface tension coefficient, as is discussed in Section 2. The coefficient $\gamma$ is connected with the local interface voltage by the Lippmann equation (1), which results from thermodynamic relations [35]. These relations are not straightforwardly included in the above mathematical model. Therefore, an examination whether the present results, obtained from the full electrohydrodynamical model, satisfy the Lippmann equation (1), would yield an independent verification. We rewrite r.h.s. of Eq. (17) for $d\gamma/dT = 0$ as

$$q_s E_x(x, y = b) = \frac{\partial \gamma}{\partial x} \tag{28}$$

from which

$$\gamma(x) = \int_0^x q_s E_x(x, y = b) dx \quad . \tag{29}$$

We use the numerical solution to calculate $\gamma(x)$ and then $\frac{\partial \gamma}{\partial V} = \frac{\partial \gamma}{\partial \phi}$, which we compare with calculated profile of $q_s$.

The result of this comparison is shown in Fig. 5 for three values of the applied voltage. We observe coincidence of all profiles to within the plot accuracy. We also observe that both compared values are scaled by the applied voltage $V_0$. This shows, in particular, that non-linear coupling of the surface charge with velocity in Eq. (12) is weak. Since the velocity and the surface charge are coupled by equations (9) and (12), and the Lippmann equation results from thermodynamic relations, this comparison yields a good verification of the numerical results.

### 7.2. Effect of Joule heating

The electric conductivity of the corn oil is of the order of $10^{-10}$ (Table 1), so that the voltage induced electric current and the Joule heating are negligible. However, when the liquid becomes slightly electrically conducting, observations change drastically. For each of the liquids 2, 3, and 4, considered below, we choose voltage, which, under effect of a fully developed flow, yields maximal temperature difference in the range of $0.5 - 2°C$.

A very rough estimate of the Joule heating effect can be obtained by considering an infinite layer of width $b$, subjected to a potential difference $V_0$ at its boundaries, so that the electric current passing through it is $j = \sigma V_0 / b$. Assuming both boundaries being isothermal at the temperature $T_0$, this leads to the temperature distribution

$$T(y) = T_0 + \frac{\sigma V_0^2}{2\kappa}\left(\frac{y}{b} - \frac{y^2}{b^2}\right). \tag{30}$$

The maximal temperature is located at $y = b/2$ and the maximal temperature difference is $T_{max} - T_0 = \frac{\sigma V_0^2}{8\kappa}$. For the three considered poorly conducting liquids this value reaches $82°C$ for the liquid 2 at $V_0 = 100V$, $267°C$ for the liquid 3 at $V_0 = 50V$, and $256°C$ for the liquid 4 at $V_0 = 1V$. These numbers show that the Joule heating cannot be neglected. It will be shown below that the above temperature differences strongly overestimate the correct result. The resulting temperature is considerably lower because of (i) heat transfer through the lateral boundaries and (ii) because of convective mixing.

Figure 6 shows temperature distribution in the lower liquid, which results from the Joule heating and not yet affected by the flow. All the boundaries are considered to be isothermal at the temperature $T_0$. The latter is withdrawn from the isotherm values shown in Fig. 6. We observe that heat losses through the lateral boundaries lead to much lower maximal temperatures than those reported above. The isotherms in all the three cases exhibit similar patterns, whose differences are caused mainly by different ratios of liquid to air thermal conductivities (see eq. (11.2)). The maximal temperature differences do not exceed $6°C$, which justifies the Boussinesq approximation.

To show how all the three electrocapillary, thermocapillary and buoyancy forces affect the flow, we perform calculations first for the electrocapillary driving only, then for combined electrocapillary and thermocapillary driving, and finally for the realistic model that accounts for all the three forces. The results are shown in Figs. 7-9.

The patterns of purely electrocapillary-driven flows (Fig. 7) are similar to those reported for a dielectric liquid flow (Fig. 2). We observe that the flow weakens with increase of the electric conductivity, meaning that electrocapillary forcing is most profound in dielectric liquids. We observe also that maximal temperatures and the isotherm patterns are similar to those reported in Fig. 6 for the no-flow case, so that convective mixing by the electrocapillary flows appears to be negligible. It should be noted here, that decrease of the flow velocity with the increase of electric conductivity observed in Fig. 7, can be caused by a stronger convection of the surface charge affected by the jump of $\sigma E_n$ in Eq. (19). Alternatively, it can be a result of a weaker production of the surface charge governed by the jump of $\varepsilon E_n$ in Eq. (4.2). The latter seems to be unlikely, since the electric permittivity of the corn oil, where the electrocapillary effect is the strongest, is smaller than that of the other liquids. The equipotential lines in all the cases considered, are similar to those shown in Fig. 3a. Inside the liquid the lines are almost vertical, so that the electric potential is a function of $x$ only, and $E_n = \partial\phi/\partial y \approx 0$. To verify this, we repeated calculation for liquid 4, increasing the electric permittivity from 5 to 70, which is the value of liquid 2. No noticeable changes in the results were observed. Thus, we conclude that larger electric conductivity leads to a stronger convection of the surface charge along the interface, so that surface charge non-uniformities are smeared and the Coulomb force decreases.

As a result of the Joule heating, the temperature maximum is located at the midline $x = l/2$ (Fig. 6). Thus, the thermocapillary force drives the flow from the center of the interface to its ends, as is seen in Fig. 8, where the thermocapillary force is combined with the electrocapillary one. One observes that the velocities become noticeably larger than those shown in Fig. 7, meaning that the thermocapillary force is much larger than the electrocapillary one. Nevertheless, the streamlines inside the liquid are not symmetric, as one would expect for a purely thermocapillary convection. At the left part of the interface direction of the two forces is the same, while at the right part it is opposite. As a result, we observe a more intensive motion and a larger convective vortex in the left part of the cavity, which also shows that interaction of the two forces breaks the flow symmetry and, therefore is not negligible. A more intensive flow leads to a stronger convective mixing, so that temperature maxima are decreased. Owing to the stronger flow in the left vortex, they are shifted towards the left boundary (cf. isotherms in Figs. 7 and 8). The more intensive flow leads to a development of temperature boundary layers near the vertical boundaries, which is expected taking into account relatively large Prandtl numbers of the three liquids (Table 2).

When the buoyancy force is added to the model (Fig. 9), the flow patterns drastically change. The maximal values of the temperature, shifted towards the interface, create an unstable stratification into the air and a stable stratification inside the liquids. Taking into account a smaller viscosity of air, these leads to development of an intensive buoyancy convection inside the air, which appears as several convective rolls. The cold air (see isotherms in Fig. 9) descends near the right boundary, so that the air motion along the interface in its right part is directed from right to left, which is opposite to what was observed in Fig. 8. Similarly to convection in horizontally elongated cavities [23,56], the air rolls rotate in the same direction and are located inside a large circulation, which moves along the interface from right to left. In the bulk of liquids, the flow is weak because it is suppressed by the stable stratification. The strongest buoyancy forces are located in thin boundary layers adjacent to the vertical boundaries, where the horizontal part of the temperature gradient is large, which leads to a large non-potential buoyancy force. One observes also the velocity boundary layers along the vertical walls, which usually develop at large Grashof and Prandtl numbers (Table 2).

In the flows shown in Fig. 9 the colder liquid temperature is located closer to the vertical walls. This means that buoyancy force there is directed downwards and purely buoyant circulations would be directed opposite to what is shown in the figure. Considering the right circulations in the streamline patterns of Fig. 9, we notice that also the electrocapillary and thermocapillary forces act in the direction opposite to the observed flow, as is shown in Fig. 8. The resulting direction of the right circulations is defined by the balance of stresses at the interface, where the air flow is so strong that the viscous interfacial stress it produces overcomes all other forces. The left circulations are not affected so strongly by the air flow, so that their direction is the same as in Fig. 8, and corresponds to the direction of all three buoyant, thermocapillary and electrocapillary forces.

To gain better comparison into the interaction of the electro- and thermocapillary forces, we plot dimensional values of their stresses in Fig. 10. The electrocapillary forces in the models with and without buoyancy are same, since the flow does not affect the volume charge (see Eq. (19)). We notice also that in the complete model with buoyancy, the thermocapillary stress can change its sign several times, which happens because convective rolls in the air affect the interfacial temperature, so that its distribution along the interface becomes non-monotonic. Comparison of the two frames of Fig. 10 shows that maximal values of the thermocapillary forces are several orders of magnitude larger than the electrocapillary ones. However, since the thermocapillary force change sign along the interface (Fig. 10b), its value in the central part is much weaker than at the ends, where the temperature gradient is larger. This makes electrocapillary and thermocapillary forces comparable at the central part of the interface, thus explaining the breaks of symmetry in Fig. 8, and showing that their interaction must be taken into account.

The above examples show that contrarily to dielectric fluids, the electrocapillary effect in weakly and strongly conducting electrolytes can be overshadowed by thermocapillary and buoyancy driving forces. A strong buoyancy force in the lighter upper phase creates flow, which is intensive enough to make the flow direction at the interface opposite to what is expected from the convective and electrocapillary driving. These results lead to a conclusion that without consideration of the two-phase model and accounting for velocity and temperature distribution into the air, it would be impossible to obtain a correct flow pattern into the lower liquid.

We are interested also to examine whether the Lippmann equation (1) holds in non-uniformly heated fluids, when their flows are driven not only by the electrocapillary force. Similarly to the isothermal case, we check whether the surface charge $q_s$ can be replaced in Eq. (17) by the derivative $d\gamma/d\phi$. Assuming that dependence of the surface tension on the electric potential and the temperature is described by independent dependencies, we again apply Eq. (29) to calculate $\gamma$, and then to calculate $d\gamma/d\phi$. The result is presented in Fig. 11. It follows that the Lippmann equation still holds and limits of its applicability are yet to be studied.

An additional observation is scaling of the surface charge and $d\gamma/d\phi$ with the applied voltage $V_0$ (Fig. 10). The two weaker conducting liquids, 2 and 3, exhibit almost the same scaled results, which are also close to that reported in Fig. 5 for a dielectric liquid. When electric conductivity increases, like in the liquid 4, the scaled results deviate from the previous ones. Thus limitations of this scaling is another issue yet to be studied.

8.  Concluding remarks

A confined setup used by Melcher and Taylor [1] for experimental demonstration of the electrocapillary flow was studied numerically. First, we considered the flow in a corn oil – air system, as in the experiment [1], for which a qualitative agreement between the calculated streamlines and the experimental flow visualization was obtained. The change of velocity pattern with the increase of the externally applied voltage was examined. It was shown that a stronger forcing leads to the intensification of flow near the boundary having lower electric potential.

An analogy between the electrocapillary and thermocapillary flows is well-known [14]. Also in this study, we observe similarity of flows driven by these two different interface forcing mechanisms and give relevant references to the thermocapillary flows results, which had been studied much more intensively than the electrocapillary ones. It raises a question of how these two mechanisms interact in a non-isothermal fluid. Noticing, that applied potential differences lead to appearance of the electric current in the liquid, we took into account the Joule heating, thus arriving to a well-defined non-isothermal model. The electric conductivity of the corn oil is extremely small (Table 1), so that the Joule heating leads to a negligible increase of the temperature, and all the above conclusions about the corn oil – air system remain valid. To make the Joule heat release more profound, and additionally motivated by recent studies of flows in LMBs [26], we replaced the corn oil by three different weakly conducting liquids that have gradually increasing electric conductivity.

Starting to study the effect of Joule heating, we noticed that heat transfer through the lateral boundaries and convective mixing must be taken into account. Otherwise, the resulting temperature will be strongly overestimated. Then we observed that with the increase of electric conductivity of the test liquid, the electrocapillary effect weakens compared to the buoyancy and thermocapillary driving.

The thermocapillary force drives the flow along the interface from the center to the lateral walls, thus producing two almost symmetric vortices. A combined action of the electrocapillary and thermocapillary forces breaks this symmetry (Fig. 8), since the electrocapillary force is unidirectional and drives the flow from the right to the left end wall. We showed that at the central part of the interface the two surface forces become comparable. As a result, the vortex adjacent to the left wall becomes stronger than another one. The isotherms in the lower liquid exhibit developing temperature boundary layers along the vertical boundaries.

In the realistic model that involves also the buoyancy forces, the flow patterns completely change. A strong buoyancy convection in the air creates large stress at the interface, which drives the flow opposite to the direction of all the three buoyancy, thermocapillary, and electrocapillary forces. From this observation, it follows that consideration of a single phase model of the lower liquid only, would lead to wrong results. Flow in the lower liquid is suppressed by the stable temperature stratification in the bulk of the cavity, while both velocity and temperature thin boundary layers develop near the vertical boundaries. Comparison of the interfacial stresses produced by the electrocapillary and thermocapillary effects shows that in the central part of the interface they are comparable, so that their interaction affects the final flow pattern.

Finally, we examine whether our numerical results satisfy the Lippmann equation, which is not a part of our model and is derived from the thermodynamic relations [35]. A good agreement obtained for isothermal (Fig. 5) and non-isothermal (Fig. 10) flows verifies present calculations.

Table 1. Properties of the considered liquids

| Liquid | | 1 | 2 | 3 | 4 | 5 |
|---|---|---|---|---|---|---|
| Property | Notation | Corn oil / source | NaCl – Water 0.1 Mole/kg solution | NaCl – Water 2 Mole/kg solution | Electrolyte LiCl-KCl | Air (at 25℃) / source |
| density, $kg/m^3$ | $\rho$ | 915. [1] | 998.8 [54,51] | 1071 [42,54] | 1598 [31] | 1.184 [42] |
| viscosity, $kg/m \cdot s$ | $\mu$ | 0.0592 [1] | 0.00103 [55] | 0.001075 [46] | 0.0022 [31] | $1.84 \times 10^{-5}$ [42] |
| electric permittivity | $\varepsilon$ | 3.35 [1] | 70 [48] | 48 [48] | 5.0 [45] | 1.0005 [42] |
| electric conductivity, $1/\Omega \cdot m$ | $\sigma$ | $0.24 \times 10^{-10}$ 40] | 0.01 [47] | 0.125 [47] | 187.1 [31] | $10^{-14}$ [43] |
| Heat conductivity, $W/m \cdot ℃$ | $\kappa$ | 0.166 [40] | 0.61 [49] | 0.586 [49] | 0.365 [31] | 0.0263 [42] |
| Heat capacity, $J/kg \cdot ℃$ | $Cp$ | 1956 [41] | 4190 [50] | 3700 [50] | 1202 [31] | 1006 [42] |
| Heat expansion, $1/℃$ | $\beta$ | $7.22 \times 10^{-4}$ [41] | $2.67 \times 10^{-4}$ [51] | $3.7 \times 10^{-4}$ [51] | $3.32 \times 10^{-4}$ [31] | 0.00338 [42] |
| Surface tension*, $N/m$ | $\gamma$ | 0.0316 [40] | 0.0728 [52] | 0.0755 [52] | 0.0165 [44] | |
| Temperature coefficient of surface tension*, $N/m \, K$ | $d\gamma/dT$ | $0.53 \times 10^{-5}$ [40] | $0.14 \times 10^{-3}$ [53] | $0.14 \times 10^{-3}$ [53] | $3.1 \times 10^{-5}$ [31] | |
| Interface electric conductivity**, $1/\Omega \cdot m^2$ | $\sigma_s$ | $10^{-10}$ | $10^{-10}$ | $10^{-10}$ | $10^{-10}$ | |
| Interface diffusion coefficient**, $m^2/s$ | $D_s$ | $10^{-09}$ | $10^{-09}$ | $10^{-09}$ | $10^{-09}$ | |
| Interface electric mobility of ions**, $m^2/V \cdot s$ | $\xi_s$ | $10^{-09}$ | $10^{-09}$ | $10^{-09}$ | $10^{-09}$ | |

* with air
** the correct value is unknown

Table 2. Dimensionless parameters and their characteristic values for the Taylor-Melcher experiment, $\Delta T = T_1 - T_0$

| Dimensionless parameter | Definition | Corn oil, $V_0 = 20 kV$ | NaCl – Water 0.1 Mole/kg solution, $V_0 = 70V$ | NaCl – Water 2 Mole/kg solution, $V_0 = 20V$ | LiCl-KCl, $V_0 = 1V$ |
|---|---|---|---|---|---|
| Prandtl number, $Pr$ | $\dfrac{\mu_1}{\alpha_1 \rho_1}$ | 697 | 7.07 | 6.79 | 7.245 |
| Grashof number, $Gr$ | $\dfrac{g\beta_1(T_1 - T_0)H^3\rho_{1,0}^2}{\mu_1^2}$ | $93\Delta T$ | $1.35\times10^5\Delta T$ | $1.98\times10^5\Delta T$ | $9.43\times10^4\Delta T$ |
| Marangoni number, $Ma$ | $\dfrac{\gamma(T_1 - T_0)b}{\mu_1\alpha_1}$ | $37\Delta T$ | $3.54\times10^4\Delta T$ | $3.35\times10^4\Delta T$ | $2818\Delta T$ |
| Electrocapillary number, $Ec$ | $\dfrac{\varepsilon_1 V_0^2 \rho_1}{\mu_1^2}$ | 193.6 | 2859 | 158 | 0.0098 |
| Joule heating number, $Jo$ | $\dfrac{\sigma V_0^2 b^2}{\mu_1 c_{p,1}(T_1 - T_0)}$ | $1.2\times10^{-7}$ | 0.0164 | 0.0182 | $2.55\times10^6$ |
| Galileo number, $Ga$ | $\dfrac{g\rho_{1,0}^2 b^3}{\mu_1^2}$ | $1.29\times10^5$ | $5.06\times10^8$ | $5.34\times10^8$ | $2.84\times10^8$ |
| Bond number, $Bo$ | $\dfrac{\rho_{1,0}g b^2}{\gamma}$ | 410 | 194 | 201 | 1372 |
| Capillary number, $Ca*$ | $\dfrac{\mu_1^{4/3}g^{1/3}}{\gamma\rho^{1/3}} = \dfrac{Bo}{Ga^{2/3}}$ | 0.16 | $3.06\times10^{-4}$ | $3.05\times10^{-4}$ | 0.0032 |
| Taylor electrocapillary number, $Ta*$ | $\dfrac{\varepsilon\rho^{1/3}V_0^2}{\mu_1^{4/3}g^{1/3}b}$ | 61.4 | 3.59 | 0.194 | $1.49\times10^{-5}$ |

* definition of Ref. [24]

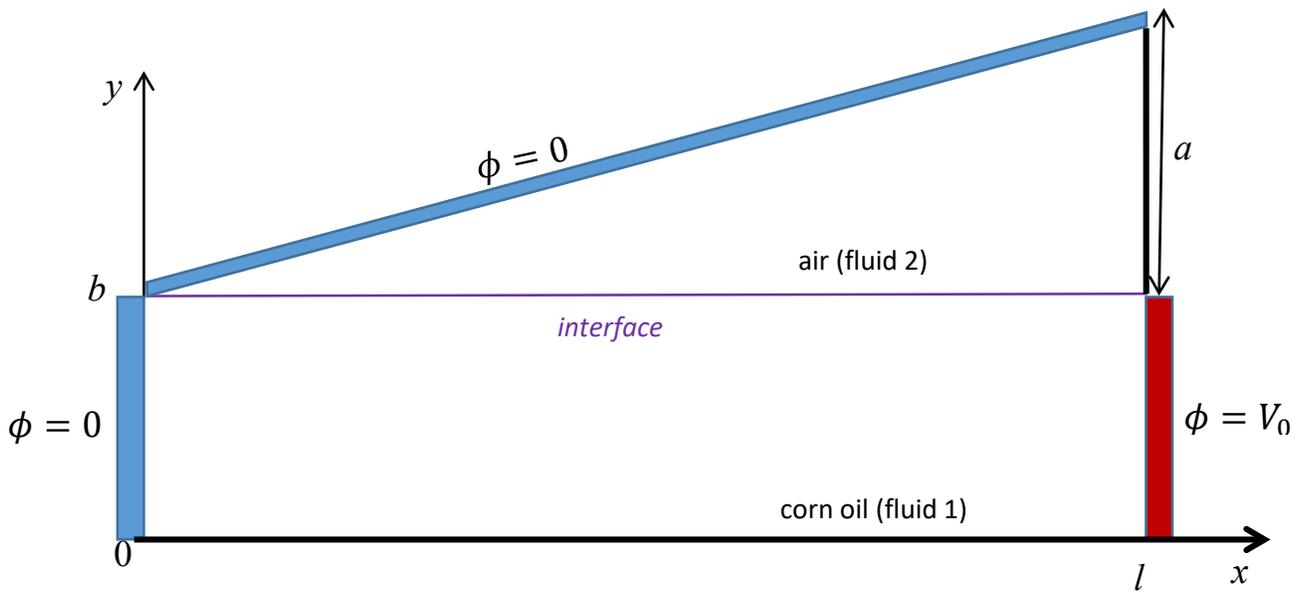

Fig. 1. Sketch of the problem

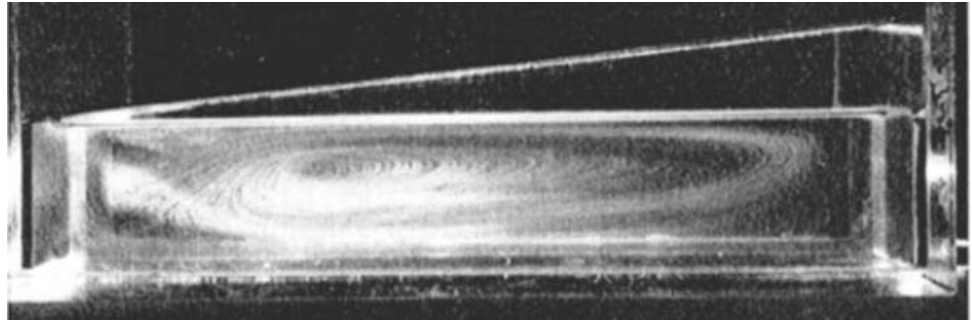

(a)

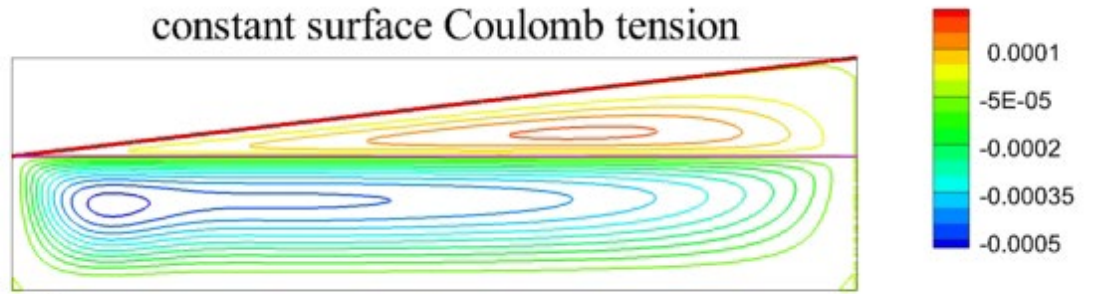

(b)

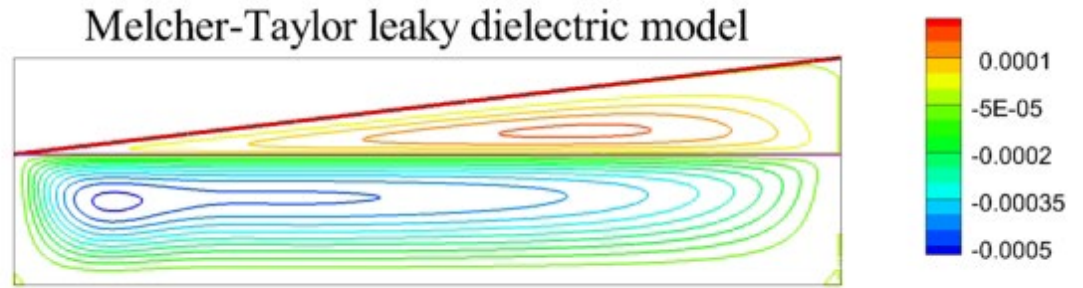

(c)

Fig. 2. (a) Flow image of the experiment [1] at $V_0 = 20\ kV$. Streamlines of the computed flow (b) with a constant surface Coulomb tension (13); $\psi_{min} = -0.000523\ m^2/s$ , $\psi_{max} = 0.000211\ m^2/s$ . (c) with a complete leaky dielectric model (1)-(11); $\psi_{min} = -0.000516\ m^2/s$ , $\psi_{max} = 0.000213\ m^2/s$. Along the interface the flow is directed from the right to the left boundary.

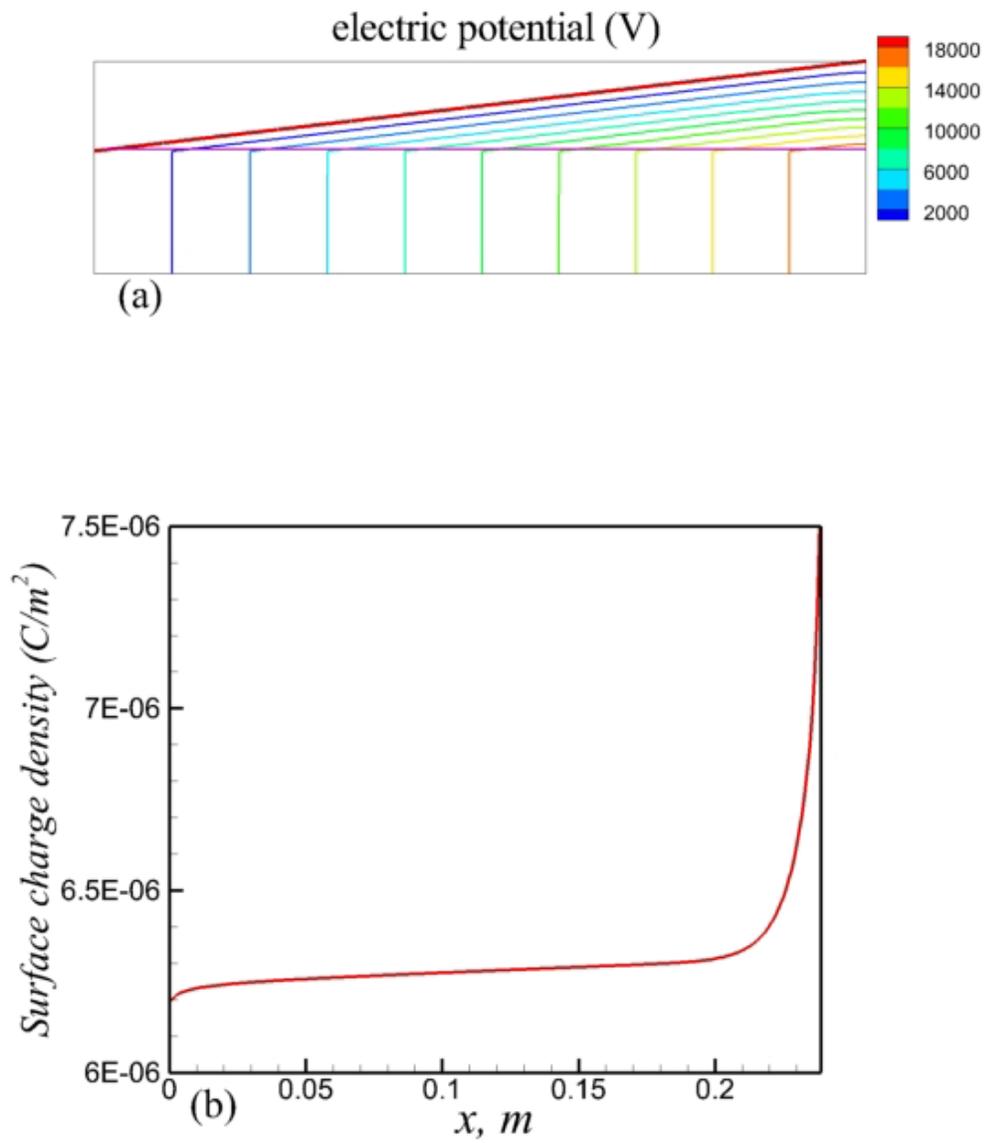

Fig. 3. Equipotential lines (a) and distribution of the surface charge density along the surface (b).

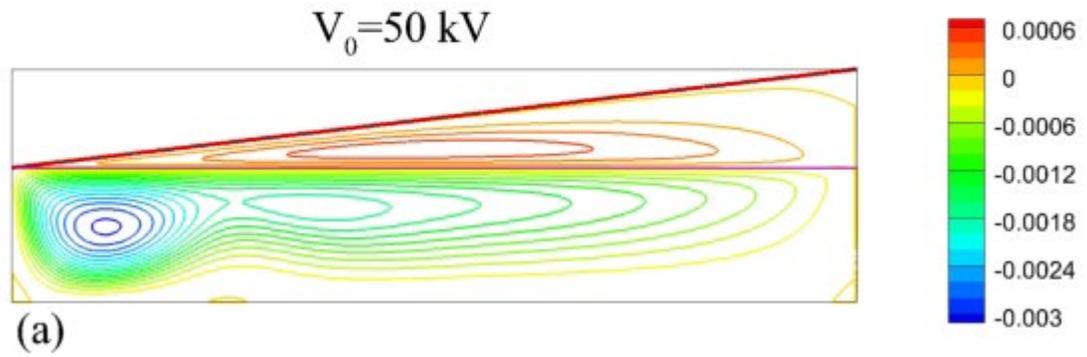

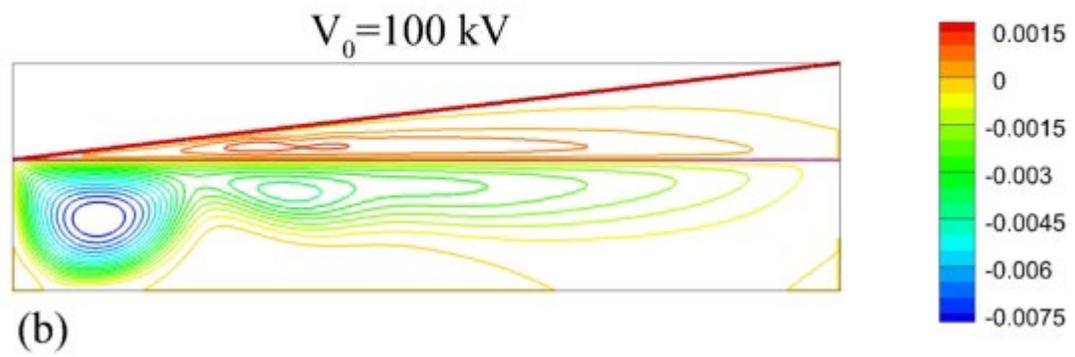

Fig. 4. Change in streamlines with the increase of applied voltage. Along the interface the flow is directed from the right to the left boundary.

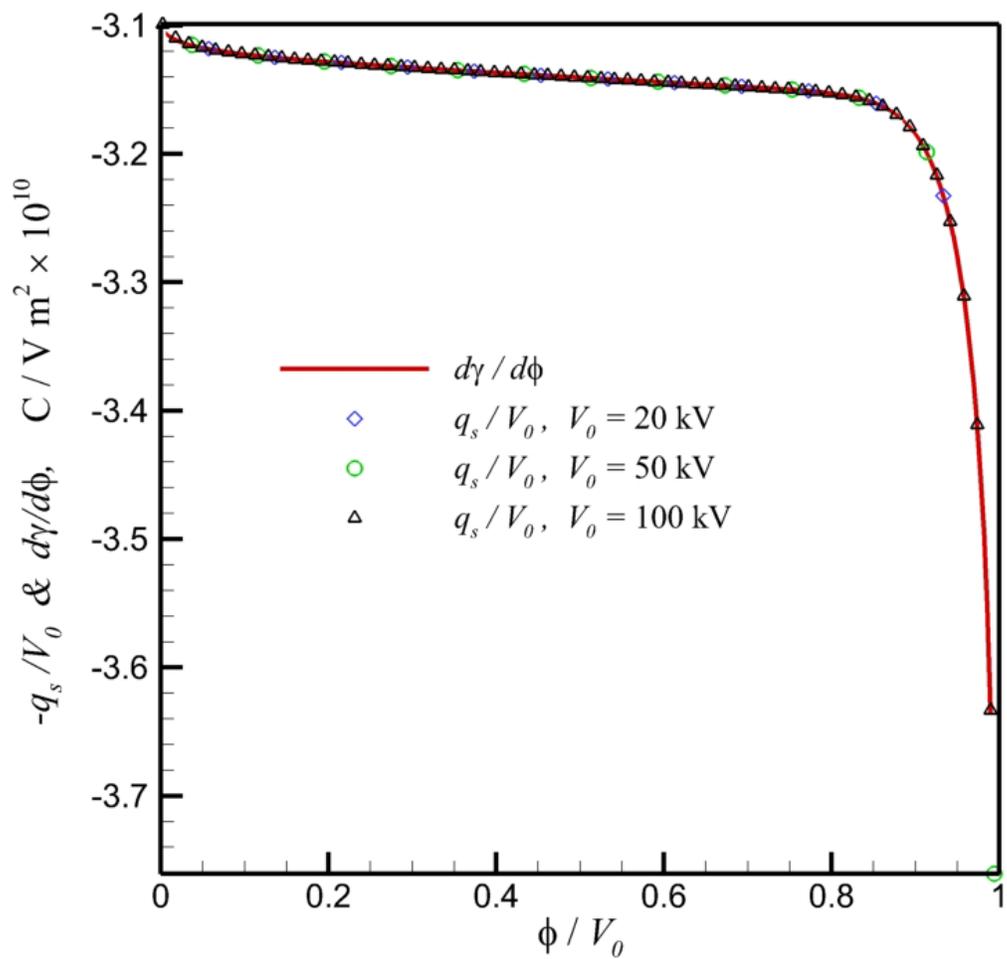

Fig. 5. Illustration of validity of the Lippmann equation: comparison of potential dependence of the surface tension and distribution of the surface charge.

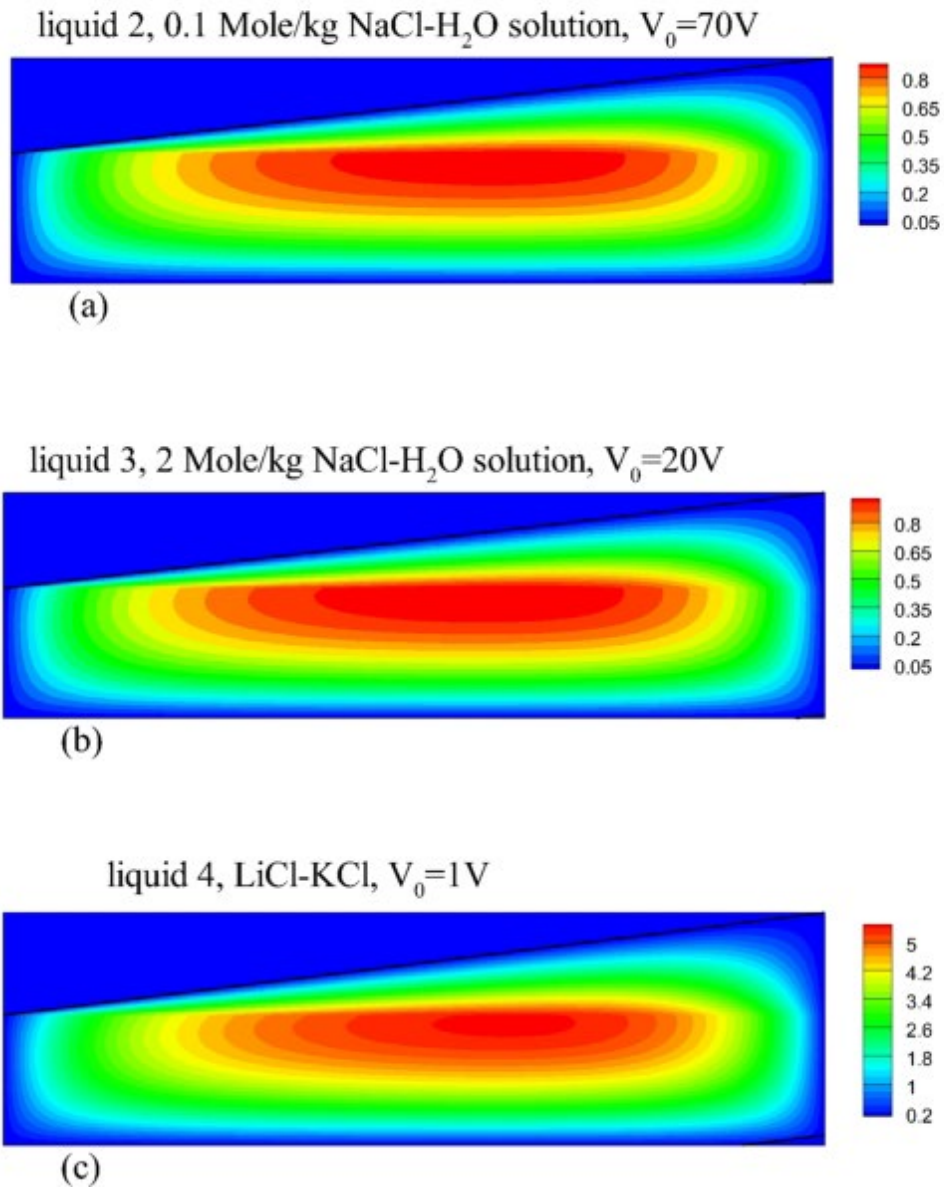

Fig. 6. Isotherms resulting by a Joule heating in the considered geometry, not affected by flow.

liquid 2, 0.1 Mole/kg NaCl-H$_2$O solution, V$_0$=100V

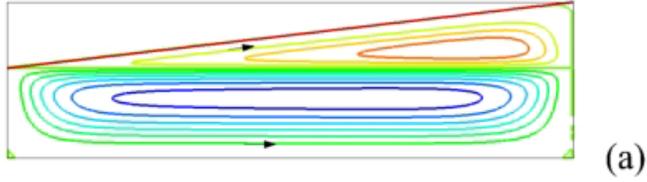
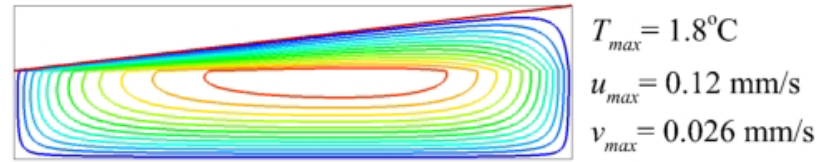

$T_{max} = 1.8^{\circ}$C
$u_{max} = 0.12$ mm/s
$v_{max} = 0.026$ mm/s

(a)

liquid 3, 2 Mole/kg NaCl-H$_2$O solution, V$_0$=50V

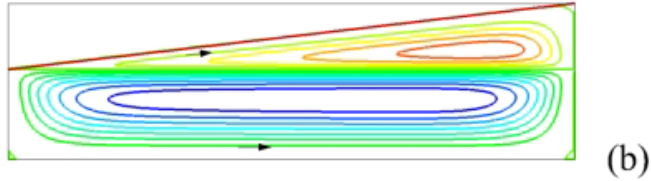
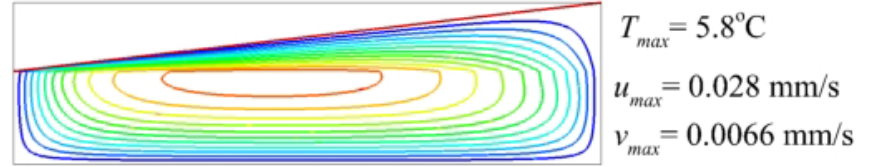

$T_{max} = 5.8^{\circ}$C
$u_{max} = 0.028$ mm/s
$v_{max} = 0.0066$ mm/s

(b)

liquid 4, LiCl-KCl, V$_0$=1V

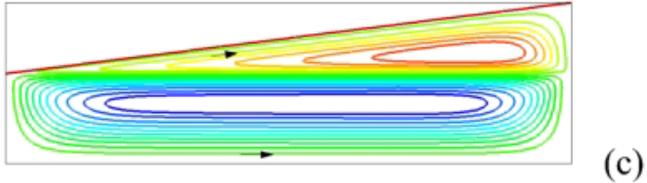
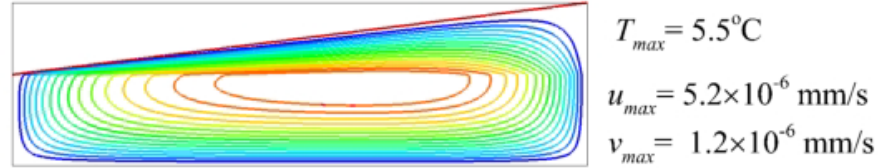

$T_{max} = 5.5^{\circ}$C
$u_{max} = 5.2 \times 10^{-6}$ mm/s
$v_{max} = 1.2 \times 10^{-6}$ mm/s

(c)

Fig. 7. Streamlines and isotherms of flows driven by the electrocapillary force only. Maximal horizontal and vertical velocity values in the liquid: (a) 0.0045 mm/s, 0.0013 mm/s; (b) $3.5 \times 10^{-3}$ mm/s, $1.1 \times 10^{-3}$ mm/s; (c) $3.4 \times 10^{-6}$ mm/s, $1.2 \times 10^{-6}$ mm/s .

liquid 2, 0.1 Mole/kg NaCl-H$_2$O solution, V$_0$=100V

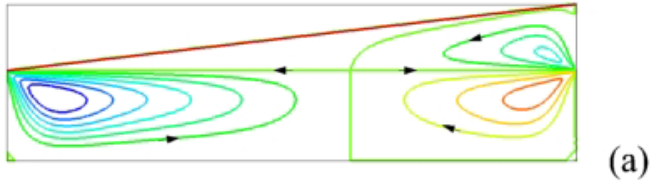
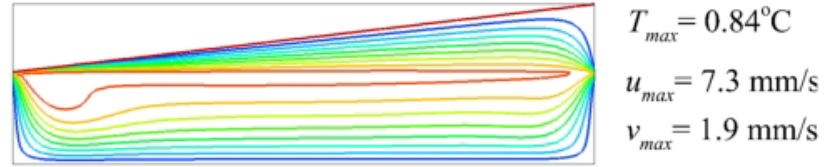

$T_{max} = 0.84\,^{\circ}\mathrm{C}$
$u_{max} = 7.3$ mm/s
$v_{max} = 1.9$ mm/s

(a)

liquid 3, 2 Mole/kg NaCl-H$_2$O solution, V$_0$=50V

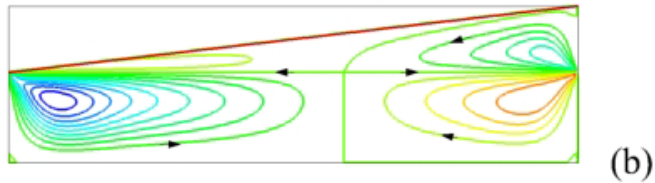
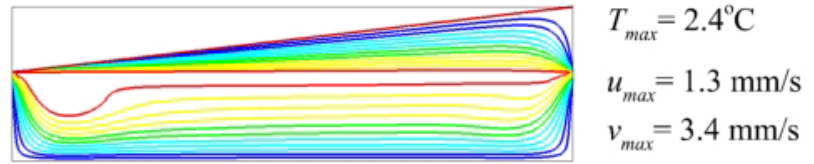

$T_{max} = 2.4\,^{\circ}\mathrm{C}$
$u_{max} = 1.3$ mm/s
$v_{max} = 3.4$ mm/s

(b)

liquid 4, LiCl-KCl, V$_0$=1V

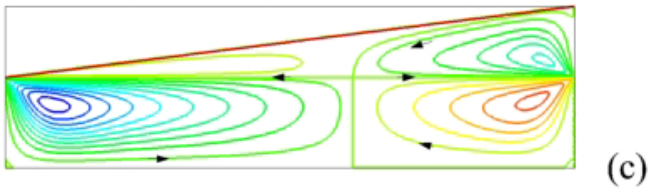
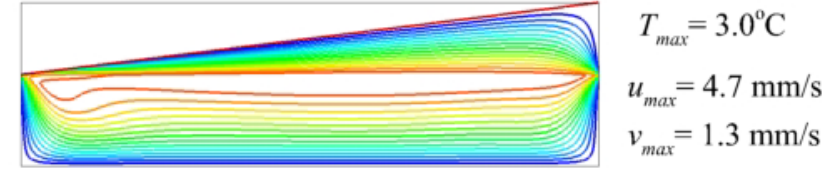

$T_{max} = 3.0\,^{\circ}\mathrm{C}$
$u_{max} = 4.7$ mm/s
$v_{max} = 1.3$ mm/s

(c)

Fig. 8. Streamlines and isotherms of flows driven by the electrocapillary and thermocapillary forces with neglected buoyancy force. Maximal horizontal and vertical velocity values in the liquid: (a) 1.53 mm/s, 0.69 mm/s; (b) 1.52 mm/s, 0.69 mm/s; (c) 1.49 mm/s, 0.59 mm/s .

liquid 2, 0.1 Mole/kg NaCl-H$_2$O solution, V$_0$=100V

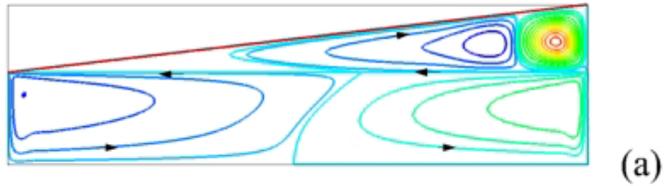 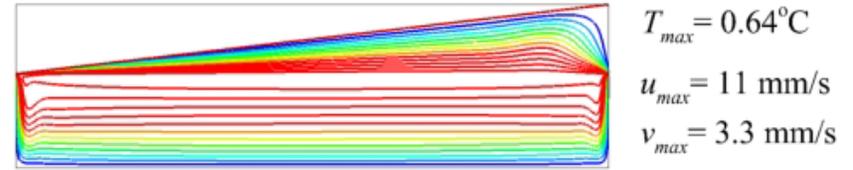

$T_{max}$ = 0.64$^o$C

$u_{max}$ = 11 mm/s

$v_{max}$ = 3.3 mm/s

(a)

liquid 3, 2 Mole/kg NaCl-H$_2$O solution, V$_0$=50V

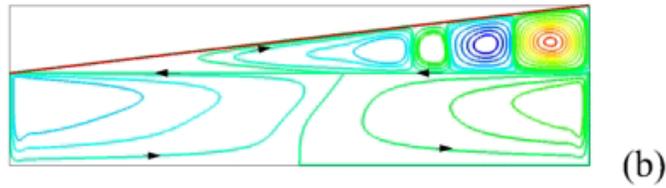 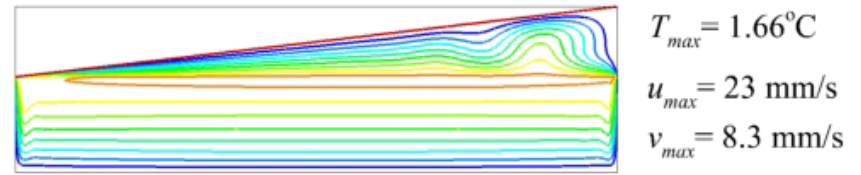

$T_{max}$ = 1.66$^o$C

$u_{max}$ = 23 mm/s

$v_{max}$ = 8.3 mm/s

(b)

liquid 4, LiCl-KCl, V$_0$=1V

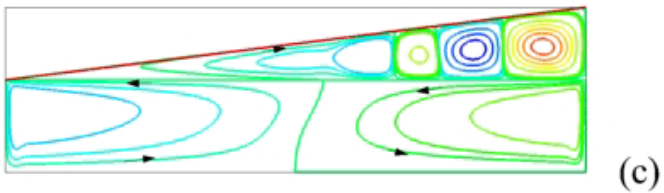 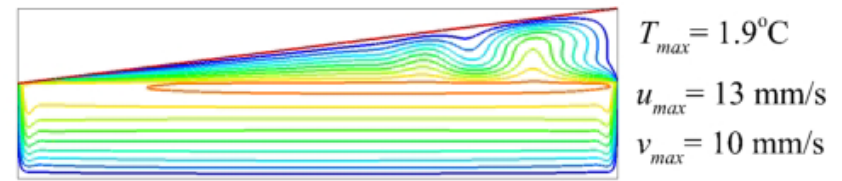

$T_{max}$ = 1.9$^o$C

$u_{max}$ = 13 mm/s

$v_{max}$ = 10 mm/s

(c)

Fig. 9. Streamlines and isotherms of flows driven by all three forces. Maximal horizontal and vertical velocity values in the liquid: (a) 1.09 mm/s, 0.97 mm/s; (b) 0.2 mm/s, 0.2 mm/s (c) 0.72 mm/s, 1.56 mm/s .

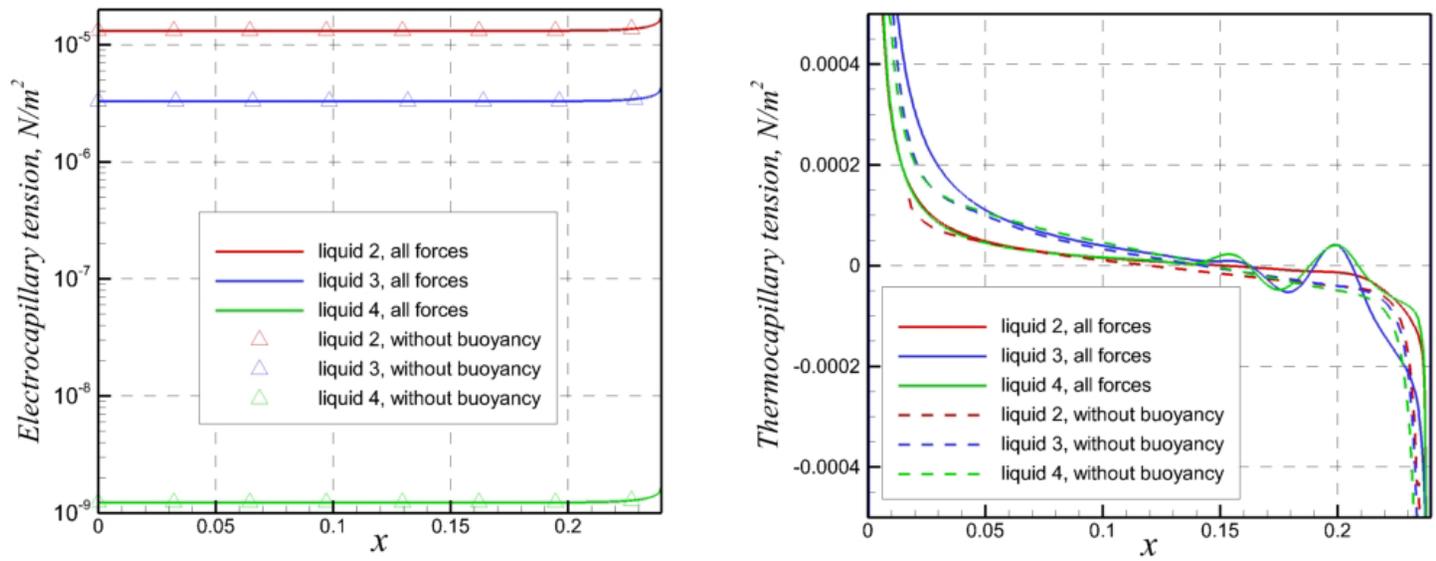

Fig. 10. Dimensional values of electrocapillary and thermocapillary forces in the different cases considered.

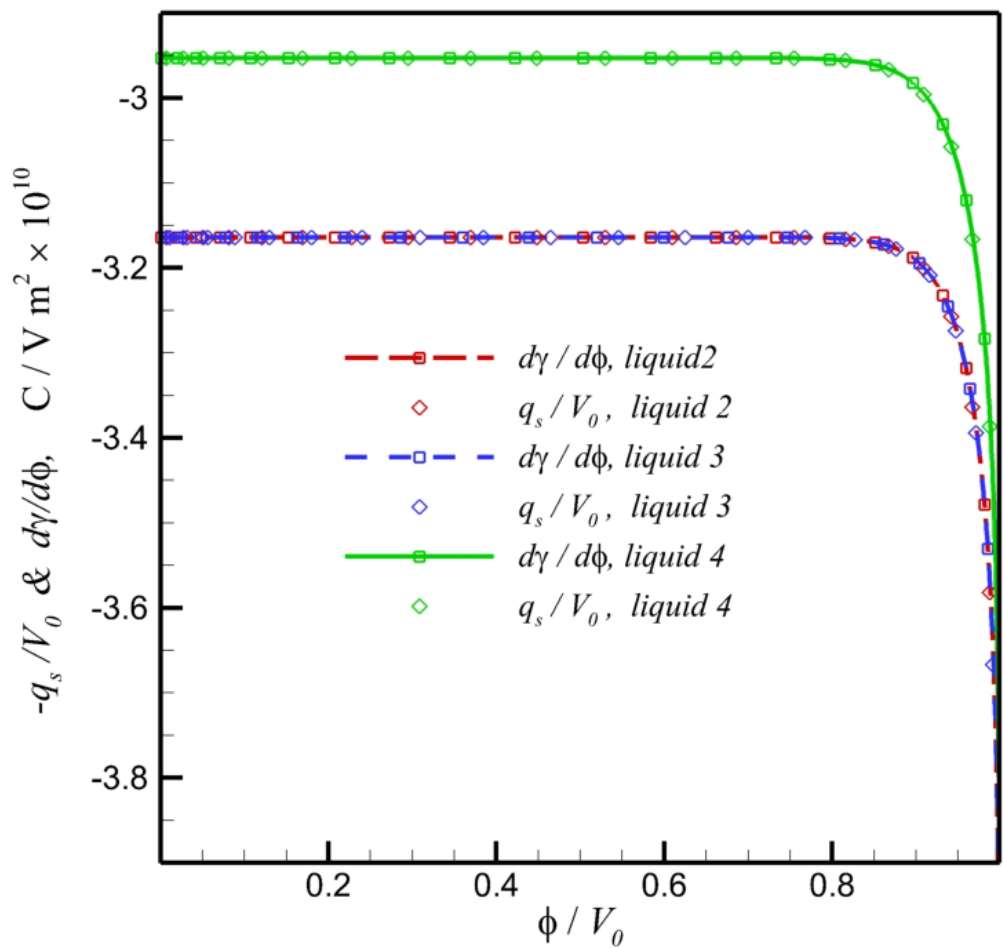

Fig. 11. Verification of the Lippmann equation for non-isothermal flows in electrolytes driven by electrocapillary, thermocapillary, and buoyancy forces: comparison of potential dependence of the surface tension and distribution of the surface charge.